\documentclass[%
 reprint,
 amsmath,amssymb,
 aps,
]{revtex4-1}

\usepackage{graphicx}
\usepackage{dcolumn}
\usepackage{bm}

\begin{document}
 \qquad \qquad \qquad \qquad \qquad \qquad \qquad CERN-TH-2017-182

\title{ Positivity of the real part of the forward scattering amplitude}

\author{Andr\'{e} Martin}
\affiliation{CERN, Geneva, Switzerland}

\author{Tai Tsun Wu}%
\affiliation{%
 Gordon McKay Laboratory, Harvard University, \\Cambridge, MA, U.~S.~A. \\
}

\date{\today}

\begin{abstract}
We prove the general theorem that the real part of the forward two-body scattering amplitude is positive at sufficiently high energies if, above a certain energy, the total cross section increases monotonically to infinity at infinite energy. 
\end{abstract}

\pacs{Valid PACS appear here}

\maketitle


\section{\label{sec:sec1} Historical Introduction}

In 1965, Khuri and Kinoshita published a series of interesting papers on the real part of the forward scattering amplitude obtained via dispersion relations from the imaginary part which is proportional to the total cross section~\cite{ref1}. Among their results was the prediction that, if the Froissart bound is saturated, i.e., if the total cross section behaves at high energies like $(\ln s)^2$ where $s$ is the square of the center-of-mass energy~\cite{ref2}, then the real part is positive at high energies. Moreover, the ratio $\rho$ of the real part to the imaginary part would behave as $\frac{\pi}{\ln s}$ at high energies. At that time, nobody thought that the total cross section would increase at high energies. Instead, the general belief was that the Froissart-Martin bound is an upper bound and that all total cross sections would approach constants or decrease to zero at high energies. Earlier, in 1960, Gribov showed that it is not possible to have at the same time a total cross section approaching a finite non-zero limit and a diffraction peak approaching a fixed shape as a function of the momentum transfer~\cite{ref3}. Instead, he preferred the total cross section to decrease to zero. In 1962, it was discovered that the diffraction peak for proton-proton scattering was shrinking~\cite{ref4}.

In view of this situation, Cheng and one of us (TTW) decided to learn about the high-energy behavior of total cross sections by studying quantum gauge field theory, specifically by summing the leading terms of the perturbation series.  The result found this way in 1970 was a surprise: the total cross section must increase at high energies, essentially saturating the Froissart-Martin bound, and the real part of the forward scattering amplitude does have the Khuri-Kinoshita behavior~\cite{ref5}. At that time, the measured proton-proton total cross section was still decreasing, while the real part of the forward scattering amplitude was still negative but increasing rapidly ~\cite{ref6}. Three years later, two experiments at the Interesting Storage Ring (ISR) at CERN showed that the proton-proton total cross section did turn around and start rising ~\cite{ref7}. At CERN, after a talk by Amaldi on these experimental results, one of us (AM) ``predicted" that the real part would become positive. Actually it was a guess at that time; a rigorous proof is to be presented in the present paper. A few years later, in 1977, an experiment at CERN showed that the real part was indeed becoming positive ~\cite{ref8}. Years later, the rise of the total cross section and the positively of the real part were both confirmed ~\cite{ref9, ref10}. 

Using the shrinking of the width of the diffraction peak and the rise of the total cross section, one of us (AM) proved in 1997 that, if the differential cross section at fixed negative $t$, for $-T<t<0$, decreases to zero sufficiently rapidly and if the total cross section increases to infinity, then the real part must change sign at least once in this interval $-T<t<0$~\cite{ref11}. Here $t$ is as usual the negative of the square of the momentum transfer and T a positive number arbitrarily small. It is in fact in re-examining this theorem, the proof of which still seems rather mysterious, that we were led to study again the problem of the real part in the forward direction $t=0$. Various results were obtained, the most striking one being the one presented in this paper. However, these considerations have failed to lead to an alternative proof of the result of reference~\cite{ref11}.

\section{\label{sec:sec2} Theorem on the real part and its proof}

The measurement of the real part of the scattering amplitude in the forward direction is of great importance. In the cases of proton-proton and prion-proton scatterings, when the measured total cross sections were still decreasing as function of energy, the first indication that this total cross section would turn around and increase came from the measurement of the real parts~\cite{ref5}. Specifically, at that time the measured values were negative but becoming less so when the center-of-mass energy increased, as expected. However, the increase (i.e., less negative) was too fast and showed a tendency to overshoot to become positive. Indeed, this was the first indication that the proton-proton total cross section, among others, would increase. As already mentioned, the theoretical  prediction of increasing total proton-proton cross section was first made in 1970 and the experimental observation in 1973. 

It is the purpose of the present paper to study in general the sign of the real part of the forward scattering amplitude. The specific problem is: under what general and realistic conditions on the total cross section, the real part of the forward two-body scattering amplitude can be guaranteed to be positive at sufficiently high energies?

Clearly, this problem should be studied through dispersion relations.  Consider the elastic scattering process

\begin{equation}
a + b \rightarrow a +b;
\label{aba:eq1}
\end{equation}

\noindent let $s,~t,~u$ be the Mandelstam variables. Throughout this paper, $t=0$ so that the scattering is in the forward direction. In this case, 

\begin{equation}
s + u = 2m_a^2 + 2m_b^2
\label{aba:eq2}
\end{equation}

\noindent where $m_a$ and $m_b$ are the masses of the particles $a$ and $b$. Thus, when $t=0$, the $s-u$ symmetric variable is 

\begin{equation}
\xi = \frac{1}{2} (s-u) = s-m_a^2 -m_b^2.
\label{aba:eq3}
\end{equation}

This scattering amplitudes for $a+b \rightarrow a+b$ and $a+\bar{b} \rightarrow a+\bar{b}$ are in general different. Define $f(\xi)$ to be the average of the forward scattering amplitude for these two processes and $\sigma(\xi)$ that of the total cross sections, then

\begin{equation}
Im ~f(\xi) = \xi \sigma(\xi)
\label{aba:eq4}
\end{equation}

\noindent approximately at high energies.  Here an unimportant multiplicative constant has been omitted.

The main result of the present paper is: 

\noindent Theorem

If, for sufficiently large values of $\xi$, $\sigma(\xi)$ is non-decreasing and approaches infinity as $\xi \rightarrow \infty$. Then $Re ~f(\xi)$ is positive for all sufficiently large values of $\xi$.

The proof begins with the dispersion  relation for $f(\xi)$

\begin{equation}
Re  ~f(\xi) -f(0) = \frac{2\xi^2}{\pi} \int_\mu^\infty  \frac{d\xi^\prime}{\xi^\prime} \frac{lm ~f(\xi)}{\xi^{\prime 2} - \xi^2}
\label{aba:eq5}
\end{equation}

\noindent where $\mu = 2m_a m_b$. Let $\xi_o$ be the value of $\xi$ such that 

\begin{equation}
\frac{d\sigma}{d\xi} \geq 0  \qquad \qquad for \qquad \qquad   \xi >\xi_0.
\label{aba:eq6}
\end{equation}

The dispersion relation (\ref{aba:eq5}) can be re-written as

\begin{equation}
\hat{f}(\xi) = \frac{\xi}{\pi} \int^\infty_{\xi_0} d\xi^\prime \sigma( \xi^\prime)  (\frac{1}{\xi^{\prime} - \xi} - \frac{1}{\xi^\prime + \xi}), 
\label{aba:eq7}
\end{equation}

\noindent where

\begin{equation}
\hat{f}(\xi) = Re ~f(\xi) -f(0) - \frac{2\xi^2}{\pi} \int^{\xi_0}_\mu d\xi^\prime \sigma(\xi^\prime) \frac{1}{\xi^{\prime 2} - \xi^2}.
\label{aba:eq8}
\end{equation}

\noindent Since the last term in (\ref{aba:eq8}) is asymptotically, for large $\xi$,

\begin{equation}
\frac{2\xi^2}{\pi} \int^{\xi_0}_\mu  d\xi^\prime \sigma(\xi^\prime) \frac{1}{\xi^{\prime 2}-\xi^2} \sim -\frac{2}{\pi} \int_\mu^{\xi_0} d\xi^\prime \sigma (\xi^\prime),
\label{aba:eq9}
\end{equation}

\noindent the difference $\hat{f}(\xi) - Re ~f(\xi)$ is bounded in absolute value. 

Integrating the right-hand side of (\ref{aba:eq7}) by parts leads to

\begin{equation}
\begin{split}
\begin{aligned}
&\text{RHS of} ~(\ref{aba:eq7}) =  \frac{\xi}{\pi} \int^\infty_{\xi_0} d\xi^\prime  \sigma(\xi^\prime) \frac{\partial}{\partial\xi^\prime}  ln  \vert \frac{\xi^\prime -\xi}{ \xi^\prime + \xi} \vert \\
&= \frac{\xi}{\pi} \sigma (\xi_0) ln \vert \frac{\xi_0+\xi}{\xi_0-\xi}\vert + I(\xi),
\label{aba:eq10}
\end{aligned}
\end{split}
\end{equation}

\noindent where

\begin{equation}
I(\xi) = \frac{\xi}{\pi} \int^\infty_{\xi_0} d\xi^\prime  \frac{d\sigma(\xi^\prime)}{d\xi^\prime} ln \vert \frac{\xi^\prime+\xi}{\xi^\prime -\xi} \vert.
\label{aba:eq11}
\end{equation}

\noindent Note that the first term on the right-hand of (\ref{aba:eq10}) is again bounded. 

It only remains to show that this $I(\xi)$ increases without bound for $\xi\rightarrow\infty$ when 

\begin{equation}
\frac{d\sigma}{d\xi} \geq 0
\label{aba:eq12}
\end{equation}

\noindent for large $\xi$ and 

\begin{equation}
\sigma(\xi) \rightarrow \infty
\label{aba:eq13}
\end{equation}

\noindent as $\xi \rightarrow \infty$. It follows from 

\begin{equation}
ln \vert \frac{\xi^\prime + \xi}{\xi^\prime -\xi}\vert \geq ln \vert \frac{\xi+\xi_0}{\xi-\xi_0}\vert \geq \frac{2\xi_0}{\xi} 
\label{aba:eq14}
\end{equation}

\noindent when $\xi >\xi^\prime$ that

\begin{equation}
\begin{split}
\begin{aligned}
&I(\xi) \geq \frac{\xi}{\pi} \int^\xi_{\xi_0}  d\xi^\prime \frac{d\sigma(\xi^\prime)}{d\xi^\prime} ln \vert \frac{\xi^\prime+\xi}{\xi^\prime-\xi} \vert \\
&\geq \frac{\xi}{\pi} \int^\xi_{\xi_0} d\xi^\prime \frac{d\sigma(\xi^\prime)}{d\xi^\prime}  \frac{2\xi_0}{\xi}\\
&= \frac{2\xi_0}{\pi} [\sigma(\xi) -\sigma(\xi_0)]
\label{aba:eq15}
\end{aligned}
\end{split}
\end{equation}

\noindent This proves that, because of (\ref{aba:eq13}), $I(\xi)$ increases without bound as $\xi\rightarrow\infty$. 

The result is therefore, with the conditions (\ref{aba:eq12}) and (\ref{aba:eq13}),

\begin{equation}
\begin{split}
\begin{aligned}
&Re ~f(\xi) \geq f(0) + \frac{2\xi^2}{\pi} \int^{\xi_0}_\mu d\xi^\prime \sigma(\xi^\prime) \frac{1}{\xi^{\prime 2-\xi^2}} \\
& -\frac{\xi}{\pi} \sigma (\xi_0) ln \vert \frac{\xi_0+\xi}{\xi_0-\xi} \vert  + \frac{2\xi_0}{\pi} [\sigma(\xi) -\sigma(\xi_0)]. 
\label{aba:eq16}
\end{aligned}
\end{split}
\end{equation}

\noindent On the right-hand side of this (\ref{aba:eq16}), the term $\frac{2\xi_0}{\pi}\sigma(\xi)$ approaches infinity as $\xi\rightarrow\infty$, which all the other terms are bounded. Therefore
\begin{equation}
Re ~f(\xi) > 0
\label{aba:eq17}
\end{equation}

\noindent for all sufficiently large values of $\xi$. 

This proves the Theorem. 

\section{\label{sec:sec3} Discussion}

Without any additional work, the lower bound for $Re~f(\xi)$ as given by (\ref{aba:eq16}) can be improved as follows. Let $\xi_1$ be any value between $\xi_0$ and $\xi$, then the $I(\xi)$ defined by (\ref{aba:eq11}) clearly satisfies 

\begin{equation}
I (\xi) \geq \frac{\xi}{\pi} \int^\xi_{\xi_1} d\xi^\prime  \frac{d\sigma(\xi^\prime)}{d\xi^\prime} ln \vert \frac{\xi^\prime+\xi}{\xi^\prime - \xi} \vert.
\label{aba:eq18}
\end{equation}

The above argument then gives, entirely similar to (\ref{aba:eq15}), 

\begin{equation}
I (\xi) \geq \frac{2\xi_1}{\pi} [\sigma(\xi)-\sigma(\xi_1)]. 
\label{aba:eq19}
\end{equation}

The improved version of  (\ref{aba:eq16}) is then

\begin{equation}
\begin{split}
\begin{aligned}
&Re ~f(\xi) \geq f(0) + \frac{2\xi^2}{\pi} \int^{\xi_0}_\mu d\xi^\prime \sigma(\xi^\prime) \frac{1}{\xi^{\prime 2}-\xi^2} \\
& -\frac{\xi}{\pi} \sigma (\xi_0) ln \vert \frac{\xi_0+\xi}{\xi_0-\xi} \vert + ^{max}_{\xi_0 \leq \xi_1 \leq \xi} \frac{2\xi_1}{\pi} [\sigma(\xi) -\sigma(\xi_1)]. 
\label{aba:eq20}
\end{aligned}
\end{split}
\end{equation}

If, for example, $\sigma(\xi)$ saturates the Froissart-Martin bound, then this maximum over $\xi_0 \leq \xi_1 \leq \xi$ is reached when $\xi_1 \sim \xi/e$.

In some cases, (\ref{aba:eq20}) leads to a considerable improvement over (\ref{aba:eq16}). As an example, if for large $\xi$

\begin{equation}
\sigma(\xi) \sim  c~(ln~\xi)^\gamma 
\label{aba:eq21}
\end{equation}

\noindent with $0<\gamma\leq2$, then (\ref{aba:eq16}) gives 
\begin{equation}
Re ~f(\xi) > const. + \frac{2}{\pi} c\xi_0 (ln\xi)^\gamma, 
\label{aba:eq22}
\end{equation}

\noindent while (\ref{aba:eq20}) gives 

\begin{equation}
Re ~f(\xi) > const. + \frac{2\gamma}{\pi e} c\xi (ln\xi)^{\gamma-1}.
\label{aba:eq23}
\end{equation}

\noindent The lower bound (\ref{aba:eq23}) is much stronger than that of (\ref{aba:eq22}). In fact, this bound (\ref{aba:eq23}) differs from the exact asymptotic formula

\begin{equation}
Re ~f(\xi) \sim  \frac{1}{2} \pi c \gamma \xi (ln \xi)^{\gamma-1}
\label{aba:eq24}
\end{equation}
\\
\noindent by only a factor of $\frac{4}{\pi e}$.

Of course, in all the bounds derived here, by a slight adjustment of the constants, such as those of (\ref{aba:eq22}) and (\ref{aba:eq23}), all the $\xi 's$ can be replaced by the Mandelstam variable $s$.

\section{\label{sec:sec3} Acknowledgments}

One of us (AM) thanks Anderson Kondi Ramida Kohara for stimulating discussions and Marek Tasevsky for making his participation at the EDS 2017 Conference in Prague possible. The other one (TTW) is grateful to the hospitality at CERN, where this work has been carried out.  We thank Shaojun Sun for typing this paper and Tullio Basaglia for help in the submission process.


\begin{references}
\bibitem{ref11} A. Martin, {\it Phys. Lett. B} {\bf 404}, 137 (1977). In this text a small correction should be made. It is not sufficient to say that $F(s,t)/s$ goes to zero for negative t. Instead, the correct condition is $(ln ~s)^{1+\epsilon} ~F(s,t)/s \rightarrow 0$.
\end{references}
\end{document}